---------------------------------------------------------------------------
############################## RevTex Copy #################################
---------------------------------------------------------------------------

\documentstyle[aps]{revtex}

\newcommand{\bel}[1]{\begin{equation}\label{#1}}
\newcommand{\be}{\begin{equation}}
\newcommand{\ee}{\end{equation}}

\newcommand{\beal}[1]{\begin{eqnarray}\label{#1}}
\newcommand{\bea}{\begin{eqnarray}}
\newcommand{\eea}{\end{eqnarray}}

\newcommand{\bean}{\begin{eqnarray*}}
\newcommand{\eean}{\end{eqnarray*}}

\newcommand{\ba}{\begin{array}}
\newcommand{\ea}{\end{array}}

\newcommand{\bab}{\begin{abstract}}
\newcommand{\eab}{\end{abstract}}

\newcommand{\bml}{\begin{mathletters}}
\newcommand{\eml}{\end{mathletters}}

\newcommand{\q}{\quad}
\newcommand{\qq}{\quad\quad}

\newcommand{\bfm}[1]{\mbox{\boldmath $#1$}}

\newcommand{\dv}{\partial}

\newcommand{\bad}[2]{\left( \begin{array}{c}{#1}\\
{#2}\end{array} \right)} 

\newcommand{\bat}[3]{\left( \begin{array}{c}{#1}\\{#2}\\
{#3}\end{array} \right)} 

\newcommand{\bamd}[4]{\left( \begin{array}{cc}{#1}&{#2}\\
{#3}&{#4}\end{array} \right)}

\newcommand{\raw}{\rightarrow}

\newcommand{\Raw}{\Rightarrow}

\newcommand{\lrw}{\leftrightarrow}

\newcommand{\ag}{\alpha}
\newcommand{\bg}{\beta}
\newcommand{\cg}{\gamma}
\newcommand{\dg}{\delta}
\newcommand{\lgg}{\lambda}
\newcommand{\sg}{\sigma}
\newcommand{\pg}{\phi}
\newcommand{\tg}{\theta}
\newcommand{\og}{\omega}

\newcommand{\bc}{\begin{center}}
\newcommand{\ec}{\end{center}}

\newcommand{\bt}{\begin{tabular}}
\newcommand{\et}{\end{tabular}}

\newcommand{\th}[1]{\thanks{#1}}

\newcommand{\f}[1]{\footnote{#1}}

\newcommand{\bref}{\begin{references}}
\newcommand{\eref}{\end{references}}
\newcommand{\bi}{\bibitem}

\newcommand{\go}{\section{Introduction}}
\newcommand{\con}{\section{Conclusions}}

\newcommand{\axp}[3]{Ann.~Phys.~(NY)                    {\bf #1},  #2  (19#3)}
\newcommand{\cxf}[3]{Commun.~Math.~Phys.                {\bf #1},  #2  (19#3)}
\newcommand{\mxb}[3]{Mod.~Phys.~Lett.~A                 {\bf #1},  #2  (19#3)}
\newcommand{\nxb}[3]{Nucl.~Phys.                        {\bf #1},  #2  (19#3)}
\newcommand{\pxf}[3]{Phys.~Rev.~D                       {\bf #1},  #2  (19#3)}
\newcommand{\pxi}[3]{Phys.~Lett.                        {\bf #1},  #2  (19#3)}
\newcommand{\pxxa}[3]{Prog.~Theor.~Phys.                {\bf #1},  #2  (19#3)}

\newcommand{\xxx}[3]{{\bf #1},  #2  (19#3)}

\title{QUATERNIONIC ELECTROWEAK THEORY}

\author{Stefano De Leo\th{Electronic mail: {\sl deleos@le.infn.it}}
 and Pietro Rotelli\th{Electronic mail: {\sl rotelli@le.infn.it}}}
\address{Dipartimento di Fisica, Universit\`a di Lecce\\
Istituto Nazionale di Fisica Nucleare, sezione di Lecce\\
Lecce, 73100, Italy}

\date{\today}

\draft

\begin{document}

\maketitle

\bab
We explicitly develop a quaternionic version of the electroweak theory, 
based on the local gauge group $U(1, \; q)_{L}\mid U(1, \; c)_{Y}$. 
The need of a complex projection for our Lagrangian and the physical 
significance of the anomalous scalar solutions are also discussed.
\eab


\renewcommand{\thefootnote}{\sharp\arabic{footnote}}

\go
Not many mathematicians can claim to have invented a new kind of number. A 
rigorous definition of the reals was given by Eudoxus, after the 
Pythagoreans discovery that the equation $x^{2}=2$ cannot be solved for 
rational numbers. The Indian mathematician Brahmagupta was the first to 
allow zero and negative numbers to be subjected to arithmetical operations, 
thus permitting the translation from ${\cal R}^{(+)}$ to $\cal R$. Cardano, 
perhaps better known as a physician than as a mathematician, introduced 
complex numbers, probably to solve equations such as $x^{2}+1=0$. After Gauss 
had proved the fundamental theorem of 
algebra, there was no longer any need to introduce new numbers to solve 
equations~\cite{lam}. In fact, it was with a different motivation in mind that 
quaternions were invented by WR Hamilton~\cite{ham}.

Hamilton was looking for numbers of the form

\bml
\be
x+iy+jz\q ,
\ee
with
\be
i^{2}=j^{2}=-1\q ,
\ee
\eml

\noindent which would do for the space what 
complex numbers had done for the plane. 
Nevertheless such a number system does not represent a right choice. 
Working with only two imaginary units we must express the product $ij$ by
\bel{two}
ij=a+ib+jc\qq (a, \; b, \; c \in {\cal R})\q . 
\ee
The eq.~(\ref{two}) implies
\[ i^{2}j=ia-b+(a+ib+jc)c = ... +jc^{2} \]
and so the inconsistent relation
\[ c^{2}=-1\q .\] 

In 1843 Hamilton introduced a third imaginary unit $k=ij$. Numbers of the 
form

\bml
\bel{three}
q=a+ib+jc+kd\qq (a, \; b, \; c, \; d \in {\cal R})\q ,
\ee
were called quaternions. They were added, subtracted and multiplied 
according to usual law of arithmetic, except for their non commutative 
multiplication law, due to the following rules for the imaginary 
units $i, \; j, \; k$
\be 
i^{2}=j^{2}=k^{2}=-1\q ,
\ee
\be
ij=-ji=k \q ,\q jk=-kj=i \q ,\q ki=-ik=j\q .
\ee
\eml

\noindent The conjugate of~(\ref{three}) is given by
\be
q^{\dag}=a-ib-jc-kd\q .
\ee
We observe that $qq^{\dag}$ and $q^{\dag}q$ are both equal to real number
\[ N(q)=a^{2}+b^{2}+c^{2}+d^{2} \; \; ,\]
which is called the norm of $q$. When $q\neq 0$, we can define
\[q^{-1}=q^{\dag}/N(q) \; \; , \]
so the quaternions form a zero-division ring. Such a non commutative  number 
field is denoted, in Hamilton's honour, by $\cal H$.

Our aim in this work is to show that quaternions can be used to express 
standard physical theories. In particular, overcoming difficulties due to 
their non-commutative multiplication law, we formulate a quaternionic 
version of the Salam-Weinberg model. In the next section we apply the 
quaternionic numbers in classical 
and quantum physics analyzing a quaternionic formulation of special 
relativity and giving a quaternionic version of the Dirac equation. In this 
section we also point out the possible predictive potentialities of 
quaternionic numbers by a qualitative study of the Schr\"odinger equation.  
A feature of our formalism, namely the need of a complex projection for 
quaternionic Lagrangians, is discussed in the third section. In the 
following section we examine the main 
differences between the standard (complex) physical theory and our 
quaternionic version (see the doubling of solutions in the quaternionic 
bosonic equations). There, recalling the main steps of a previous article, we 
give a possible interpretation for 
the anomalous solutions which appear in the quaternionic Klein-Gordon equation. 
In the fifth section, we will explicitly discuss a Quaternionic electroweak 
Theory (QewT), based on the ``quaternionic Glashow group''
\[ U(1, \; q)_{L}\mid U(1, \; c)_{Y}
\qq (L \lrw \mbox{left-handed helicity}\; , \; 
Y \lrw \mbox{weak-ypercharge})\q . \]
Our conclusions are drawn in the final section. In the next sections we 
will adopt the following notation
\bean
\pg \;  , \; \psi \q &\lrw& \q \mbox{complex fields}\q ,\\
\Phi \; , \; \Psi \q &\lrw& \q \mbox{quaternionic fields}\q ,
\eean
and use the system of natural units ($\hbar=c=1$).

\section{Quaternions in classical and quantum physics}
If we represent complex numbers in a plane by 
\[x+iy=re^{i\tg} \]
(recall that in place of $i$ we could use any imaginary unit), 
we immediately observe that a rotation of $\ag$-angle around the $z$-axis, 
can be given by $e^{i(\tg +\ag)}$, in fact
\bel{plane}
e^{i\ag} (x+iy)=re^{i(\tg+\ag)}\q .
\ee
Using quaternions (instead of complex numbers) we can express a rotation in 
three-dimensional space. For example a rotation about an axis passing 
through the origin and parallel to a given unitary vector 
${\bf\hat u}\equiv(u_{x}, \; u_{y}, \; u_{z})$ by an angle $\ag$, can be 
obtained by making the transformation
\bel{space}
e^{(iu_{x}+ju_{y}+ku_{z})\frac{\ag}{2}}(ix+jy+kz)
e^{-(iu_{x}+ju_{y}+ku_{z})\frac{\ag}{2}}\q ,
\ee
upon the position vector ${\cal X}\equiv ix+jy+kz$. This reduces, 
after simple manipulations, to~(\ref{plane}) [pose 
$u_{x}=u_{y}=0$ and $u_{z}=1$ in~(\ref{space})], except for the appearance 
of $k$ instead of $i$. The previous 
transformation leaves ${\cal X}^{\dag}{\cal X}=x^{2}+y^{2}+z^{2}$ invariant.

The special theory of relativity requires the invariance of the expression 
\[ t^{2}-x^{2}-y^{2}-z^{2} \] 
under a coordinate transformation passing from a 
stationary frame to a moving one with constant velocity. This 
suggests that space and time be joined together in a quaternionic 
number
\be
{\cal X}\equiv t+ix+jy+kz \; \; ,
\ee
with the Lorentz invariant $Re \; {\cal X}^{2}$. If complex 
numbers are the natural candidates to represent rotation in a 
plane, quaternions express concisely the Lorentz transformations. Introducing  
{\em barred-operator} $A\mid b$, which acts on quaternionic objects $q$ as 
in
\[ (A\mid b) q=Aqb \q ,\] 
we can quickly express the generators of the Lorentz groups by the following 
operators~\cite{del1}
\bean
\mbox{boost} \q (t, \; x) \q & 
\q \frac{k \; \vert \; j - j \; \vert \; k}{2} \q ,\\
\mbox{boost} \q (t, \; y) \q & 
\q \frac{i \; \vert \; k - k \; \vert \; i}{2} \q ,\\
\mbox{boost} \q (t, \; z) \q & 
\q \frac{j \; \vert \; i - i \; \vert \; j}{2} \q ,\\
\mbox{rotation around $x$} \q & 
\qq  \frac{i-1 \; \vert \; i}{2} \q ,\\
\mbox{rotation around $y$} \q & 
\qq  \frac{j-1 \; \vert \; j}{2} \q ,\\
\mbox{rotation around z} \q & 
\qq  \frac{k-1 \; \vert \; k}{2} \q .
\eean
The last three reproduce in a new form eq.~(\ref{space}).

An interesting application of quaternions in quantum physics is represented 
by the quaternionic formulation of the Dirac equation~\cite{rot}. 
Notwithstanding the 
two-component structure of the wave function, all four standard solutions 
appear. This represents a stimulating example of the doubling of solutions 
within a quaternionic quantum mechanics with complex geometry. We indicate 
with the terminology complex geometry the use of a complex scalar product
$<\Psi\mid \Phi>_{c} \; $, defined in terms of the quaternionic counterpart 
$<\Psi\mid \Phi>_{c}$ by
\be
<\Psi\mid \Phi>_{c} \; =\frac{1-i\mid i}{2} <\Psi\mid \Phi> \; \; .
\ee
Such a scalar product was used by Horwitz and Biedenharn in order to 
define consistently multiparticle quaternionic states~\cite{hor}.

We justify the choice of a complex geometry by recalling that although there 
is in quaternionic quantum mechanics an anti-self-adjoint operator, 
$\bfm{\dv}$, with all 
the properties of a translation operator, imposing a 
quaternionic geometry, there is no corresponding quaternionic 
self-adjoint operator 
with all the properties expected for a momentum operator. This hopeless 
situation is also highlighted in Adler's recent book~\cite{adl}.  
Nevertheless we can overcome such a difficulty using a complex scalar 
product and defining as the appropriate momentum operator
\be
{\bf p}\equiv -\bfm{\dv}\mid i \q ,
\ee
note that the usual choice ${\bf p}\equiv -i\bfm{\dv}$, still gives a 
self-adjoint operator with standard commutation relations 
with the coordinates, but such an operator does not commute with the 
Hamiltonian, which will be, in general, a quaternionic quantity. Obviously, 
in order to write equations 
relativistically covariant, we must treat the 
space components and time in the same way, hence we are obliged to 
modify the standard equations by the following substitution
\be
i\dv_{t} \raw \dv_{t} \mid i \q ,
\ee
and so the quaternionic Dirac equation becomes
\be
\dv_{t} \Psi i = ( \bfm{\ag}\cdot {\bf p} + \bg m) \Psi 
\qq ({\bf p}\equiv-\bfm{\dv}\mid i) \q .
\ee
Noting that the Dirac algebra upon the {\em reals} (but not upon complex) 
has a two dimensional irreducible representation with quaternions. Thus the 
standard $4\times 4$ complex matrices ($\bfm{\ag}, \bg$) reduce to $2\times 2$ 
quaternionic matrices. For example, a particular representation is given by 
\[\bg=\bamd{1}{0}{0}{-1} \q , \q 
\bfm{\ag} = {\bf Q} \bamd{0}{1}{-1}{0}
\qq [{\bf Q}\equiv(i, \; j, \; k)]\q .\]
In this representation the quaternionic plane wave solutions are
\beal{dirac}
E>0  & : & u \; e^{-ipx} \; \; , \; \;  u \; je^{-ipx} \; \; , \nonumber \\
     &   &  \\
E<0  & :  & v \; e^{-ipx} \; \; , \; \; v \; je^{-ipx} \; \; , \nonumber
\eea
where
\[u=\sqrt{E+m}\bad{1}{-\frac{{\bf Q}\cdot {\bf p}}{E+m}} \q ,\]
and
\[v=\sqrt{\vert E\vert +m}
\bad{-\frac{{\bf Q}\cdot {\bf p}}{\vert E \vert+m}}{1} \q .\]
Following the standard approach we can define the hermitian spin 
operator 
\[ {\bf S} \equiv -\frac{{\bf Q}\mid i}{2} \q ,\]
so the four complex orthogonal solutions given in~(\ref{dirac}) correspond 
to positive and negative energy solutions with $S=\frac{1}{2}$, and for 
${\bf p}=(p_{x}, \; 0, \; 0)$, $S_{x}=\frac{1}{2}, \; -\frac{1}{2}, \; 
\frac{1}{2}, \; -\frac{1}{2},$ respectively. Thus, although our wave 
function has a two component structure, we find the four standard solutions 
to the Dirac equation. This is a {\em desirable} example of the 
so-called {\em doubling of solutions}.

Obviously such a doubling of solutions occurs also in the nonrelativistic 
Schr\"odinger equation
\be 
\dv_{t} \Psi i = - \frac{\bfm{\dv}^{\; 2}}{2m} \Psi 
\qq (\mbox{quaternionic solutions : } \; e^{-ipx}, \; je^{-ipx})\q .
\ee
If Schr\"odinger had worked within a quaternionic quantum mechanics with 
complex geometry, finding two complex-orthogonal solutions to his 
equation, he would have probably discovered {\em spin}. Indeed the non 
relativistic limit of the Dirac equation yields the Schr\"odinger-Pauli 
equation with two solutions which is formally identical with the 
{\em one-component} Schr\"odinger equation with quaternions. We like to call 
this stimulating situation within quaternionic quantum mechanics with complex 
geometry: {\em The belated theoretical discovery of spin}. 

As we have already noted elsewhere~\cite{del2} this doubling 
of solutions in the Schr\"odinger equation would be an 
impressive argument in favor of the use of quaternions within quantum 
mechanics, if this doubling of solutions did not occur also in bosonic 
equations, where it has obviously nothing to do with spin. For example, 
we find four complex orthogonal solutions for the Klein-Gordon equation, 
with the result that, in addition to the two normal solutions
\[ e^{-ipx}\qq (\mbox{positive and negative energy})\q ,\]
we discover two {\em anomalous} solutions
\[je^{-ipx}\qq (\mbox{positive and negative energy})\q .\]
The physical significance of the anomalous solutions has been a ``puzzle'' 
for the authors. Only recently, by a quaternionic study of the electroweak 
Higgs sector, we have been able to identify anomalous Higgs 
particles~\cite{del3}.

\section{Complex projected Lagrangians}
Before analyzing the Higgs sector within a QewT, 
we must highlight a feature of quaternionic field theory, namely {\em the 
need of using a complex projection for our quaternionic Lagrangians}. This 
result has been justified in previous papers~\cite{del3,del4}, here we 
briefly recall only some of the main steps.

The standard free Lagrangian density for two hermitian scalar fields 
$\pg_{1}$, $\pg_{2}$ is
\bel{2}
{\cal L}=\frac{1}{2} \; (\dv_{\mu}\pg_{1}\dv^{\mu}\pg_{1}-
m^{2}\pg_{1}^{\; 2}+\dv_{\mu}\pg_{2}\dv^{\mu}\pg_{2}-
m^{2}\pg_{2}^{\; 2}) \q ,
\ee
where
\[\pg_{1, \; 2}=V^{-\frac{1}{2}}\sum_{{\bf k}} 
(2\og_{{\bf k}})^{-\frac{1}{2}} [a_{1, \; 2}({\bf k}) e^{-ikx}+
a_{1, \; 2}^{\dag}({\bf k}) e^{+ikx}]\q .\]
The Lagrangian~(\ref{2}) can be concisely rewritten, by complex scalar 
fields $\pg$, $\pg^{\dag}$, as follows
\be
{\cal L}=\dv_{\mu}\pg^{\dag}\dv^{\mu}\pg-
m^{2}\pg^{\dag} \pg \qq (\pg \equiv \frac{\pg_{1} + i \pg_{2}}{\sqrt{2}} \; ,  
\; \pg^{\dag} \equiv \frac{\pg_{1} - i \pg_{2}}{\sqrt{2}})\q .
\ee
Note that the cross term $\pg_{1}i\pg_{2}-i\pg_{2}\pg_{1}$ is trivially 
null ($[\pg_{1}, \; \pg_{2}]=0$ and $i$ commutes with $\pg_{1, \; 2}$). 
We wish now to extend the previous considerations to four hermitian scalar 
fields $\pg_{1}$, $\pg_{2}$, $\pg_{3}$, $\pg_{4}$. In order to rewrite
\bea
{\cal L} & = & \frac{1}{2} \; (\dv_{\mu}\pg_{1}\dv^{\mu}\pg_{1}-
m^{2}\pg_{1}^{\; 2}+\dv_{\mu}\pg_{2}\dv^{\mu}\pg_{2}-
m^{2}\pg_{2}^{\; 2} \; + \nonumber \\
         &   & + \; \dv_{\mu}\pg_{3}\dv^{\mu}\pg_{3}-
m^{2}\pg_{3}^{\; 2}+\dv_{\mu}\pg_{4}\dv^{\mu}\pg_{4}-
m^{2}\pg_{4}^{\; 2}) 
\eea
by quaternionic scalar fields $\Phi$, $\Phi^{\dag}$,
\be
\Phi \equiv \frac{\pg_{1} + i \pg_{2}+ j \pg_{3}+ k \pg_{4}}{\sqrt{2}} \qq 
(\Phi^{\dag} \equiv \frac{\pg_{1}-\pg_{2}i-\pg_{3}j-\pg_{4}k}{\sqrt{2}}) \q ,
\ee
we must require a complex projection of our Lagrangian. Such a 
complex projection kills the ``pure'' quaternionic cross terms
\[ \pg_{1}j\pg_{3}-\pg_{3}j\pg_{1} \q , \q  
\pg_{1}k\pg_{4}-\pg_{4}k\pg_{1} \q , \q 
- \; \pg_{2}ij\pg_{3}-\pg_{3}ji\pg_{2} \q , \q 
- \; \pg_{2}ik\pg_{4}-\pg_{4}ki\pg_{2} \q .\]
So the quaternionic Klein-Gordon Lagrangian, with four hermitian scalar fields, 
reads\f{In order to maintain the canonical commutation 
relations for the creation-annihilation operators, we must assume either 
commutation or anticommutation relations with $j$
\bean
[a({\bf k}), \; a^{\dag}({\bf k'})]=\dg_{{\bf k}{\bf k'}} &  
\Raw & -j[a({\bf k}), \; a^{\dag}({\bf k'})]j=-j\dg_{{\bf k}{\bf k'}}j \\
 & \Raw & [-ja({\bf k})j, \; -ja^{\dag}({\bf k'})j]=\dg_{{\bf k}{\bf k'}} \\
 & \Raw & -ja({\bf k})j= \pm a({\bf k}) \; \; .
\eean}
\bel{sc}
{\cal L}_{c}=\frac{1-i\mid i}{2} \; (\dv_{\mu}\Phi^{\dag}\dv^{\mu}\Phi-
m^{2}\Phi^{\dag} \Phi) \; \; .
\ee
The complex projection for scalar Lagrangian can be avoided by requiring 
``double-barred'' operators within quaternionic scalar fields\f{For 
{\em double-barred} operators, $a\parallel e^{\pm ikx}$, the 
exponential acts from the right on the vacuum state (or any state vector), 
whereas for {\em barred} operators, 
$A\mid b$, $b$ acts from the right on the {\em fields}. If $j$ commutes with 
$a_{m}({\bf k})$ the cross terms are automatically killed.}. However 
{\em this} solution fails for fermionic fields. 

Let us consider the standard (complex) Dirac Lagrangian
\be
{\cal L}^{D}=\bar{\psi}\cg^{\mu}\dv_{\mu}\psi i -m\bar{\psi}\psi \q .
\ee
As well known, it represents an hermitian operator, in fact
\[ (\bar{\psi}\cg^{\mu}\dv_{\mu}\psi i)^{\dag}=
-i(\dv_{\mu}\bar{\psi})\cg^{\mu}\psi \q ,\]
and after integration by parts, gives (using the fact that here the fields 
are complex)
\[  i\bar{\psi}\cg^{\mu}\dv_{\mu}\psi\equiv \bar{\psi}\cg^{\mu}\dv_{\mu}\psi i 
\q .\]
In our quaternionic formalism the different position of the imaginary unit 
suggests the modification of the kinetic term in the Dirac Lagrangian. In 
order to obtain an hermitian operator we consider
\bel{ho}
\frac{1}{2} \; [\bar{\Psi}\cg^{\mu}\dv_{\mu}\Psi i-
i(\dv_{\mu}\bar{\Psi})\cg^{\mu}\Psi] \q ,
\ee
which, after integration by parts, reduces to
\[ \frac{1-i\mid i}{2} \; (\bar{\Psi}\cg^{\mu}\dv_{\mu}\Psi i) \q . \]
So a first modification of the standard Dirac Lagrangian is justified by 
the simple requirement that ${\cal L}^{D}$ be hermitian. Nevertheless this 
requirement says nothing about the Dirac mass term. It is here that we must 
invoke the ``quaternionic'' variational principle which generalizes the 
variational rule that says that $\Psi$ and $\bar{\Psi}$ must be varied 
independently.

If we consider the kinetic term~(\ref{ho}) we immediately note that a 
variation $\dg \Psi$ gives
\[\frac{1}{2} \; [\bar{\Psi}\cg^{\mu}\dv_{\mu} \dg \Psi i-
i(\dv_{\mu}\bar{\Psi})\cg^{\mu}\dg \Psi] \q ,
\]
and since within a quaternionic field theory $[\dg \Psi, \; i]\neq 0$, we 
cannot extract mechanically the field equation from the Lagrangian
\be
{\cal L}^{D} =  (\bar{\Psi}\cg^{\mu}\dv_{\mu}\Psi i)_{c} - m \bar{\Psi}\Psi
 \q .
\ee
In order to obtain the desired Dirac equation for $\Psi$ and $\bar{\Psi}$ 
we are obliged to treat $\Psi$ and $\Psi i$ (similarly $\bar{\Psi}$ and 
$i\bar{\Psi}$) as independent fields and modify the mass term in the Dirac 
Lagrangian into
\[ - \; \frac{m}{2} \; (\bar{\Psi}\Psi-i\bar{\Psi}\Psi i) \q .\]
The final result is the need of a ``full'' complex projection 
\be
{\cal L}_{c}^{D} =  \frac{1-i\mid i}{2} \; 
(\bar{\Psi}\cg^{\mu}\dv_{\mu}\Psi i - m \bar{\Psi}\Psi)
 \q .
\ee
Any complex projection, under extreme right or left multiplication by a 
complex number, behaves as follows,
\[ (z{\cal L}\tilde{z})_{c}=z{\cal L}_{c} \tilde{z}=z\tilde{z} {\cal L}_{c} 
\q .\]
Thus if $z\tilde{z}=1$ we have invariance. When the transformation is 
attributed to the fields $\Phi$ and $\Psi$, this implies that 
$\tilde{z}=z^{*}$ and hence
\[ z\in U(1, \; c) \q .\]
{\em The automatic appearance of this complex unitary group is expected 
whatever the left acting (quaternionic) unitary group is}~\cite{del3}.

If we analyze the scalar field Lagrangian~(\ref{sc}) we immediately note that 
the ``full'' quaternionic quaternionic gauge group is
\bel{gla} 
U(1, \; q)\mid U(1, \; c) \q ,
\ee
which is isomorphic at the Lie algebra level with the (complex) Glashow group 
$SU(2, \; c)\times U(1, \; c)$. This invariance group and the four 
quaternionic Klein-Gordon solutions, equal to the Higgs particles number 
before spontaneous symmetry breaking, suggests that the Salam-Weinberg theory 
contains an interpretation of the anomalous particles. 

\section{Anomalous solutions.}
Since the only fundamental scalar could be the Higgs boson, in order to 
interpret the anomalous scalars we believe to be natural to concentrate our 
attention on the Higgs sector of the electroweak theory. Moreover, as we 
pointed out in the previous section, the number of Higgs particles, before 
spontaneous symmetry breaking, is four
\[ h^{0}\q , \q  h^{-} \q , \q \bar{h}^{0} \q , \q h^{+} \q ,\]
and this agrees with the number of quaternionic solutions to the Klein-Gordon 
equation.

Remembering that the standard (complex) term $\Phi^{\dag}\Phi$ splits into 
$(\Phi^{\dag}\Phi)_{c}$ when $\Phi$ becomes a quaternionic scalar field, 
we write the quaternionic Higgs Lagrangian  as follows
\bel{hig}
{\cal L}^{H} = (\dv_{\mu}\Phi^{\dag}\dv^{\mu}\Phi)_{c}-
\mu^{2}(\Phi^{\dag} \Phi)_{c}-
\vert \lgg \vert (\Phi^{\dag} \Phi)_{c}^{\; 2} \; \; ,
\ee
with
\[\Phi\equiv h^{0} +j h^{+} \q , \q h^{0} \; \mbox{and} \;  
h^{+} \; \mbox{complex scalar fields} \q .\]
The Lagrangian~(\ref{hig}) is obviously invariant
under the global group
\[ U(1, \; q)\mid U(1, \;c) \qq [\Phi \; \raw \; \exp(-g 
{\bf Q}\cdot \bfm{\ag}/2) \; \Phi \; 
\exp (i \tilde{g} Y_{\Phi} \bg /2) ] \q . \]
If we wish impose a local gauge invariance we must compensate the 
derivative terms which appear in the Lagrangian by introducing a quaternionic 
covariant derivative
\bel{li}
\dv^{\mu}\Phi \raw {\cal D}^{\mu}\Phi \equiv \dv^{\mu}\Phi-\frac{g}{2} \; 
(i\Phi W_{1}^{\mu}+j\Phi W_{2}^{\mu}+k\Phi W_{3}^{\mu}) + \frac{\tilde{g}}{2} 
\; Y_{\Phi} \Phi B^{\mu} i \q .
\ee
The hermitian gauge fields ${\bf W}^{\mu}$ and $B^{\mu}$ have the well known 
gauge transformation properties
\bc
\bt{lcl}
${\bf W}^{\mu}$ & $\raw$ & ${\bf W}^{\mu}-\dv^{\mu}\bfm{\ag}-g \bfm{\ag} \wedge
{\bf W}^{\mu} \; \; ,$ \\
$B^{\mu}$ & $\raw$ & $B^{\mu}-\dv^{\mu}\bg \; \; .$
\et
\ec
Thus the Higgs Lagrangian, invariant under the {\em local} group 
$U(1, \; q)\mid U(1, \; c)$, reads
\bel{higli}
{\cal L}^{H} = [({\cal D}_{\mu}\Phi)^{\dag}{\cal D}^{\mu}\Phi]_{c}-
\mu^{2}(\Phi^{\dag} \Phi)_{c}-
\vert \lgg \vert (\Phi^{\dag} \Phi)_{c}^{\; 2} \; \; .
\ee
Let us consider $\mu^{2}<0$ and examine the consequences of spontaneous 
symmetry breaking. As in the standard theory, we choose a {\em real} minimum 
value for the Higgs potential
\[ \Phi_{0} \; = \frac{v}{\sqrt{2}} 
\qq (v=\sqrt{-\mu^{2} / \vert \lgg \vert } \; ) \q ,\]
which breaks both $U(1, \; q)$ and $U(1, \; c)$ symmetries, but preserve an 
invariance under the symmetry generated by a {\em mixed} generator
\bc
\bt{lclc}
                         &     & $\q -i\Phi_{0} \;=-i\frac{v}{\sqrt{2}} \q $
& , \\
 $U(1, \; q)$ generators &  :  & $\q -j\Phi_{0} \; =-j\frac{v}{\sqrt{2}} \q $
& , \\
   &     & $\q -k\Phi_{0} \; =-k\frac{v}{\sqrt{2}} \q $ & ; \\
 & & & \\
 $U(1, \; c)$  generator $(Y_{\Phi}=+1)$  &  :  & 
$\q +\Phi_{0}i=+i\frac{v}{\sqrt{2}} \q $ & ;\\
 & & &  
\et
\ec
\be
(-i +1 \mid i) \; \Phi_{0} \; =0 \q .
\ee
We can identify this residual ``complex'' group as the electromagnetic gauge 
group. Rewriting $W_{1}^{\mu}$ and $B^{\mu}$ 
as a linear combination of the physical fields $A^{\mu}$ and $Z^{\mu}$

\bml
\beal{linear}
W_{1}^{\mu} & = & \sin \tg_{w} A^{\mu} + \cos \tg_{w} Z^{\mu} \q ,\\
B^{\mu}     & = & \cos \tg_{w} A^{\mu} - \sin \tg_{w} Z^{\mu} \q , 
\eea
\eml

\noindent with $\tg_{w}$ the Weinberg angle, and
\[g \sin \tg_{w} = \tilde{g} \cos \tg_{w} = e  \q , \]
we can quickly obtain the minimal coupling in terms of the electromagnetic 
field $A^{\mu}$
\be
\dv^{\mu} \raw \dv^{\mu} - \frac{e}{2} A^{\mu} (i-1\mid i) + ... \qq
[ \; p^{\mu} \raw p^{\mu} - \frac{e}{2} A^{\mu} (1+i\mid i) + ... \; ]\q . 
\ee
The electric charge operator
\be
\frac{e}{2} \; (1 + i\mid i) \q ,
\ee
allow us to connect the complex scalar fields $h^{0}$ with the neutral Higgs 
bosons and the (anomalous) {\em pure} quaternionic scalar fields $jh^{+}$ 
with the charged Higgs bosons. 

\section{QewT}
In this section, we summarize the structure of the quaternionic electroweak 
Lagrangian. Having introduced in the previous sections the gauge group 
$U(1, \; q)\mid U(1, \; c)$, which represents the quaternionic counterpart 
of the ``complex'' Glashow group $SU(2, \; c)\times U(1, \; c)$, we wish to 
construct a fermionic Lagrangian invariant under such a group. If we 
consider a single particle (two component) field $\Psi$, we have no hope to 
achieve this. In fact the most general transformation
\[ \Psi \raw f \Psi g \qq (f, \; g \; \mbox{quaternionic numbers} ) \q , \]
is right-limited from the complex projection of our Lagrangian and 
left-limited 
from the presence of quaternionic (two dimensional) $\cg^{\mu}$ matrices. 
So we could 
only write a Lagrangian invariant under a right-acting complex $U(1, \; c)$ 
group. The situation drastically changes if we use a ``left-real'' (four 
component) Dirac 
equation, in this case we could commute the quaternionic phase and restore 
the invariance under the left-acting quaternionic unitary group\f{The 
left (right) action of the quaternionic (complex) unitary group has nothing 
to do with the helicity indices $L$ and $R$ of the Salam-Weinberg theory}. 
Indeed in this case $\Psi$ represents two fermions.

Recalling the standard representation for the $\cg^{\mu}$ 
matrices~\cite{itz} 
\[\tilde{\cg}^{0}=\bamd{1}{0}{0}{-1} \q , \q  
\bfm{\cg}=\bamd{0}{\bfm{\sg}}{-\bfm{\sg}}{0} \q ,\]
we can immediately write the desired ``left-real'' Dirac equation
\[(\tilde{\cg}^{\mu}\dv_{\mu}\mid i-m) \Psi = 0 \; \; ,\]
where
\[ \tilde{\cg}^{\mu} \equiv \cg^{\mu} \mbox{-matrices with $i$-factors 
substituted by} \; 1\mid i \q .\]
The massless fermionic Lagrangian in our QewT 
reads\f{Note that once more the complex projection kills the 
{\em undesired} cross terms. Recall that complex projection is also 
justified by the requirements of hermiticity.}
\bel{fer}
{\cal L}_{c}^{F} = (\bar{\Psi}_{l} \tilde{\cg}^{\mu}\dv_{\mu} \Psi_{l} i +
\bar{\Psi}_{q} \tilde{\cg}^{\mu}\dv_{\mu} \Psi_{q} i )_{c} \; \; ,
\ee
with
\[ \Psi_{l}=e+j\nu \q , \q \Psi_{q}=d+ju \qq (e, \; \nu, \; d, \; u \; 
\mbox{complex fermionic fields}) \q .\]
This Lagrangian is globally invariant under the following transformations:\\

\noindent 
\bt{lrclc} 
-- left-handed fermions --~~~~~~~~~~~~~ &
$\q e_{L}+j\nu_{L}$ & $\q \raw \q$ & $e^{-\frac{g}{2} {\bf Q}\cdot \bfm{\ag}} 
\; (e_{L}+j\nu_{L}) \; e^{\frac{\tilde{g}}{2} i Y_{l}^{(L)} \bg} \q$ & ,\\
 & & & &  \\
 & $\q d_{L}+ju_{L}$ & $\q \raw \q$ & $e^{-\frac{g}{2} {\bf Q}\cdot \bfm{\ag}} 
\; (d_{L}+ju_{L}) \; e^{\frac{\tilde{g}}{2} i Y_{q}^{(L)} \bg} \q$ & ,\\
 & & & &  \\
-- right-handed fermions --~~~~~~~~~~~~~ &
$\q e_{R}$ & $\q \raw \q$ & $e_{R} \; e^{\frac{\tilde{g}}{2} i Y_{e}^{(R)} \bg}$ 
& , \\
 & & & &  \\
 & $\q d_{R}+ju_{R}$  & $\q \raw \q$ & $d_{R} \; e^{\frac{\tilde{g}}{2} i Y_{d}^{(R)} \bg} 
+ j u_{R} \; e^{\frac{\tilde{g}}{2} i Y_{u}^{(R)} \bg}$ & .\\
 & & & &  
\et
\\

\noindent Requiring that the electric charge operator 
be represented (in units of $e$) 
by
\[ {\cal Q} = \frac{Y + i\mid i}{2}  \; \; ,\]
leads to the {\em weak-hypercharge} assignments
\[ Y_{l}^{(L)}=-1 \q , \q Y_{q}^{(L)}=+\frac{1}{3} \q , \q 
Y_{e}^{(R)}=-2 \q , \q Y_{d}^{(R)}=-\frac{2}{3} \q , \q 
Y_{u}^{(R)}=+\frac{4}{3} \q .\]
In order to construct a local-invariant theory, we must introduce the gauge 
bosons
\bc
\bt{cclc}
${\bf W}^{\mu}$ & for & $U(1, \; q)_{L}$ & ,\\
$ B^{\mu}$ & for & $U(1, \; c)_{Y}$ & ,
\et
\ec
by the covariant derivatives
\bc
\bt{lcl}
${\cal D}^{\mu} (e_{L}+j\nu_{L})$ & $\equiv$ & $[\dv^{\mu}-
\frac{g}{2} \; (i \mid W_{1}^{\mu}+ 
j \mid W_{2}^{\mu}+ 
k \mid W_{3}^{\mu})
- \frac{\tilde{g}}{2} \mid B^{\mu} i \; ](e_{L}+j\nu_{L}) \q ,$\\
& & \\
${\cal D}^{\mu} (d_{L}+ju_{L})$ & $\equiv$ & $[\dv^{\mu} -
\frac{g}{2} \; (i \mid W_{1}^{\mu}+
j\mid W_{2}^{\mu}+
k\mid W_{3}^{\mu}) 
+ \frac{\tilde{g}}{6}  \mid B^{\mu} i \; ](d_{L}+ju_{L}) \q ,$\\
& & \\
${\cal D}^{\mu} u_{R}$ & $\equiv$ & $(\dv^{\mu}
+ \frac{2\tilde{g}}{3} \mid B^{\mu}i) u_{R}  \q ,$\\
& & \\
${\cal D}^{\mu} d_{R}$ & $\equiv$ & $(\dv^{\mu} 
- \frac{\tilde{g}}{3} \mid B^{\mu} i) d_{R}  \q ,$\\
& & \\
${\cal D}^{\mu} e_{R}$ & $\equiv$ & $(\dv^{\mu} 
- {\tilde{g}} \mid B^{\mu} i)e_{R}  \q ,$\\
 & &
\et
\ec
\bc
- the substitution $\dv^{\mu} \raw {\cal D}^{\mu}$ in~(\ref{fer}) 
makes our Lagrangian locally invariant - .
\ec
Working with quaternions we can concisely 
express the four gauge fields ${\bf W}^{\mu}$ and $B^{\mu}$ by only one 
quaternionic gauge field. Actually, in analogy with the quaternionic Higgs 
scalar $\Phi=h^{0}+jh^{+}$ ($h^{0}$, $h^{+}$ complex scalar 
fields), we introduce the following quaternionic gauge field
\bel{qgf}
W_{\mu}=W_{\mu}^{0}+jW_{\mu}^{+} 
\qq [ \; W_{\mu}^{0}=(B_{\mu}^{0}+iW_{\mu}^{1})/\sqrt{2} 
\; \; , \; \; W_{\mu}^{+}=(W_{\mu}^{2}-iW_{\mu}^{3})/\sqrt{2} \; ] \q , 
\ee
and so the gauge-kinetic term is represented by
\bel{gau}
{\cal L}^{G}_{c}=- \; \frac{1}{2} \; (F_{\mu \nu}^{\dag}F^{\mu \nu})_{c} \q ,
\ee
with
\[ F^{\mu \nu}=\dv^{\mu} W^{\nu} - \dv^{\nu} W^{\mu} - 
g {\bf Q}\cdot {\bf W}^{\mu} \wedge {\bf W}^{\nu} \q .\] 
Now we can add an interaction term which involves Yukawa couplings of the 
scalars to the fermions,
\bel{Y}
{\cal L}_{c}^{Y} = - \; \{G_{e}\bar{e}_{R}[\Phi^{\dag}(e_{L}+j\nu_{L}) 
]_{c}+
G_{d}\bar{d}_{R}[\Phi^{\dag}(d_{L}+ju_{L})]_{c} \; + \; 
G_{u}\bar{u}_{R}[\tilde{\Phi}^{\dag}(d_{L}+ju_{L})]_{c}\} \; +  \; \; h.c. 
\qq (\tilde{\Phi}\equiv \Phi j) \q .
\ee
Eq.~(\ref{Y}) transforms under local $U(1, \; q)_{L}\mid U(1, \; c)_{Y}$ as
\bc
\bt{lcl}
$\bar{e}_{R}[\Phi^{\dag}(e_{L}+j\nu_{L})]_{c}$         & $\q \raw \q$ & 
$e^{-\frac{\tilde{g}}{2} i Y_{e}^{(R)} \bg} \; \bar{e}_{R} \;  
e^{-\frac{\tilde{g}}{2} i Y_{\Phi} \bg} \; [\Phi^{\dag}(e_{L}+j\nu_{L}) ]_{c} \;
e^{\frac{\tilde{g}}{2} i Y_{l}^{(L)} \bg} \q , $\\
 & & \\
$\bar{d}_{R}[\Phi^{\dag}(d_{L}+ju_{L}) ]_{c}$         & $\q \raw \q$ & 
$ e^{-\frac{\tilde{g}}{2} i Y_{d}^{(R)} \bg} \; \bar{d}_{R} \; 
e^{-\frac{\tilde{g}}{2} i Y_{\Phi} \bg} \; [\Phi^{\dag}(d_{L}+ju_{L})]_{c} \;
e^{\frac{\tilde{g}}{2} i Y_{q}^{(L)} \bg} \q ,$\\
 & & \\
$ \bar{u}_{R}[\tilde{\Phi}^{\dag}(d_{L}+ju_{L})]_{c}$ & $\q \raw \q$ &
$ e^{-\frac{\tilde{g}}{2} i Y_{u}^{(R)} \bg} \; \bar{u}_{R} \;
e^{+\frac{\tilde{g}}{2} i Y_{\Phi} \bg} \; [\tilde{\Phi}^{\dag}(d_{L}+ju_{L})
]_{c} \; 
e^{\frac{\tilde{g}}{2} i Y_{q}^{(L)} \bg} \q .$\\
 & & 
\et
\ec
Because the $\Psi_{R}$ fields are complex all the {\em complex} 
phase factors can be brought together. Thus invariance follows if
\[ Y_{e}^{(R)}+Y_{\Phi}-Y_{l}^{(L)}=0 \q , \q \mbox{etc.}\]
If we expand the Lagrangian about 
the minimum of the Higgs potential by writing
\be
\Phi=e^{-\frac{{\bf Q}\cdot {\bf u}}{2v}} \; (v+H^{0})/ \sqrt{2} \qq
(H^{0} \; \mbox{hermitian scalar field}) \q ,
\ee
and transforming at once to a U-gauge:
\bean
\Phi & \q \raw \q & \Phi'=e^{\frac{{\bf Q}\cdot {\bf u}}{2v}} \; \pg=
(v+H^{0})/ \sqrt{2} \q ,\\
\Psi_{l, \; q} & \q \raw \q & \Psi_{l, \; q}'=
e^{\frac{{\bf Q}\cdot {\bf u}}{2v}} \; \Psi_{l, \; q} \q ,\\
{\bf Q}\cdot {\bf W}_{\mu} & \q \raw \q & {\bf Q}\cdot {\bf W}_{\mu}' \q ,
\eean
we can reexpress our Lagrangian in terms of the U-gauge fields. The Yukawa 
term becomes
\bean
{\cal L}_{c}^{Y} & = & - \; \frac{1}{\sqrt{2}} \;  
\{G_{e}\bar{e}_{R}[(v+H^{0})
(e_{L}+j\nu_{L})]_{c}+
G_{d}\bar{d}_{R}[(v+H^{0})(d_{L}+ju_{L})]_{c} \; +\\
 & & + \; G_{u}\bar{u}_{R}[-j(v+H^{0})(d_{L}+ju_{L})]_{c}\} \;  + \; \; h.c. \\
 & = & - \; \frac{v}{\sqrt{2}} \; (G_{e}\bar{e}e+G_{d}\bar{d}d+G_{u}\bar{u}u) 
\; + \; \; \mbox{coupling between $H^{0}$ and fermions} \q ,
\eean
so the electron and the quarks $d, \; u$ acquire a mass
\[ m_{e, \; d, \; u}=G_{e, \; d, \; u} \; \frac{v}{\sqrt{2}} \; \; .\]
From the scalar term in the Lagrangian we recognize a physical Higgs 
boson with mass 
\[ M_{H}=\sqrt{-2\mu^{2}} \q ,\]
and the intermediate boson masses
\[ M_{W^{\pm}}=g \; \frac{v}{2} \q , \q 
M_{Z}=M_{W}\sqrt{1+\frac{\tilde{g}^{2}}{g^{2}}} \q .\]
The interactions among the gauge bosons and fermions may be read off from 
\bean
{\cal L}_{c}^{I} & = & - \; \frac{g}{2} \; [(\bar{e}_{L}-\bar{\nu}_{L}j) 
\tilde{\cg}_{\mu} (i\mid W_{1}^{\mu} i + j\mid W_{2}^{\mu} i + 
k\mid W_{3}^{\mu} i) (e_{L}+j\nu_{L})]_{c} \; +\\
                &  & + \; \frac{\tilde{g}}{2} \; [(\bar{e}_{L}-\bar{\nu}_{L}j) 
\tilde{\cg}_{\mu} (e_{L}+j\nu_{L}) B_{\mu}]_{c} \; +\\ 
 & & - \; \frac{g}{2} \; [(\bar{d}_{L}-\bar{u}_{L}j) 
\tilde{\cg}_{\mu} (i\mid W_{1}^{\mu} i + j\mid W_{2}^{\mu} i + 
k\mid W_{3}^{\mu} i) (d_{L}+ju_{L})]_{c} \; +\\
                &  & - \; \frac{\tilde{g}}{6} \; [(\bar{d}_{L}-\bar{u}_{L}j) 
\tilde{\cg}_{\mu} (d_{L}+ju_{L}) B_{\mu}]_{c} \; +\\ 
                  &  & - \; \frac{2\tilde{g}}{3}  \; \bar{u}_{R} \tilde{\cg}^{\mu}
u_{R} B_{\mu}   + \frac{\tilde{g}}{3} \; \bar{d}_{R} \tilde{\cg}^{\mu}
d_{R} B_{\mu}   + \tilde{g} \; \bar{e}_{R} \tilde{\cg}^{\mu}
e_{R} B_{\mu}  \q .
\eean
For the charged gauge bosons we find
\bean
{\cal L}_{c}^{W \; - \; l, \; q} & = & - \; \frac{g}{2} \; 
[(\bar{e}_{L}-\bar{\nu}_{L}j) 
\tilde{\cg}_{\mu} (j\mid W_{2}^{\mu} i + 
k\mid W_{3}^{\mu} i) (e_{L}+j\nu_{L})]_{c} \; +\\
 & & - \; \frac{g}{2} \; [(\bar{d}_{L}-\bar{u}_{L}j) 
\tilde{\cg}_{\mu} (j\mid W_{2}^{\mu} i + 
k\mid W_{3}^{\mu} i) (d_{L}+ju_{L})]_{c}\\
                                 & = &  - \; \frac{g}{\sqrt{2}} \; 
( \bar{\nu}_{L} \tilde{\cg}^{\mu} e_{L} W_{\mu}^{+} i +
\bar{u}_{L} \tilde{\cg}^{\mu} d_{L} W_{\mu}^{+} i )  \; + \; \; h.c. 
\eean
Similarly, rewriting $W_{1}^{\mu}$ and $B^{\mu}$ as a linear combination of 
the physical fields $A^{\mu}$ and $Z^{\mu}$ [see eq.~(\ref{linear}-b)], we can 
reproduce the standard results for the neutral 
gauge boson couplings to fermions.

\con
We have formulated in the previous sections a quaternionic version of the 
electroweak theory which reproduces the standard results. Notwithstanding 
the quaternionic nature of the fields, our Lagrangians are complex projected
and this represents a {\em desirable} feature of our formalism. We have 
identified the quaternionic counterpart of the complex Glashow group 
$SU(2, \; c)\times U(1, \; c)$ with $U(1, \; q)\mid U(1, \; c)$ and argued 
that the right-acting $U(1, \; c)$ group (at first sight unnatural in 
the context of quaternionic groups) is a direct consequence of the complex 
projection of our Lagrangians. Such a complex projection opens the 
door to all possible right-acting complex groups, for example if we consider 
the following fermionic fields
\[ \Psi_{q}=\bat{d_{r}+ju_{r}}{d_{g}+ju_{g}}{d_{b}+ju_{b}} \q , 
\q \Psi_{l}=e+j\nu_{e} \qq [ \; (r, \; g, \; b)\lrw (\mbox{red}, \; 
\mbox{green}, \; \mbox{blue}) \; ] \q ,
\]
we can quickly write a Lagrangian density,
\[
{\cal L}_{c}^{F} = (\bar{\Psi}_{l} \tilde{\cg}^{\mu}\dv_{\mu} \Psi_{l} i +
\bar{\Psi}_{q} \tilde{\cg}^{\mu}\dv_{\mu} \Psi_{q} i )_{c} \; \; ,
\]
invariant under the global gauge group
\[ SU(3, \; \tilde{c})\times U(1, \; q)\times U(1, \; \tilde{c}) \qq
(\tilde{c}=a +b\mid i \; \mbox{with} \; a, \; b \in {\cal R}) \q .
\]
So the complex projection of our Lagrangian, required in order to obtain 
the proper field equations, represents a fundamental ingredient in 
reformulating quaternionic electroweak theory and standard model. The 
complex projection of $\cal L$ allows us to confront our quaternionic 
Lagrangian densities with those of the standard theory by means of 
the {\em rules of translation}~\cite{del5}, 
obtained for complex scalar products. We have 
not however been able to {\em derive} a complex geometry from the 
assumption of the complex projection of $\cal L$ (such a connection is not 
yet clear to us and is currently under investigation). 

It is also important to recall that the possibility of rewriting standard 
particle physics theories in quaternionic form is a non trivial objective, 
in fact the non-commutative nature of 
quaternions alters the conventional approach (as in tensor products, 
variational calculus, bosonic equations).

We observe that our long standing perplexity upon the physical significance of 
the anomalous solutions is overcome. We had already observed that if the 
anomalous photon existed the field had to be non-hermitian and 
hence treated in analogy with the weak charged currents~\cite{del2}. 
Now our quaternionic version of the Salam-Weinberg model shows 
that the anomalous photon can be identified with one of the 
charged $W$ particles, and not with $Z^{0}$ as in our original 
hypothesis~\cite{del2}. Of course, without spontaneous symmetry 
breaking this identification would 
appear embarrassing since we would have expected the anomalous photon to 
have zero mass. In a similar manner the anomalous solutions of the Klein-Gordon 
equation for the Higgs fields have been identified in this work with the 
charged scalar fields before spontaneous symmetry breaking.

Now let us discuss the potential generalizations which the use of 
quaternions suggest. We have noted that 
$U(1, \; q)$ is the most natural quaternionic invariance 
group for particle physics and this coincides nicely with the practical 
importance of $SU(2, \; c)$ (spin, isospin, etc.). This type of argument 
(based on groups) for an underlying quaternionic number system is not as 
ephemeral as the above example seems to be. For example, while we have 
already noted that no 
complex group can be a priori excluded, the existence of quite simple 
invariance quaternionic groups such as $U(n, \; q)$, isomorphic to the 
unitary symplectic complex groups $USp(2n, \; c)$~\cite{gil}, would surely 
not go unnoticed. In short certain unusual groups could become ``natural'' with 
quaternions. A possible application of these consideration is to grand 
unification theories.

The relevance of unitary quaternionic groups and of the, as yet unexplored, 
mathematics of groups consisting of generalized quaternionic 
elements,
\[ q_{c}=p+q\mid i \q  \mbox{and} \q  q_{r}=p+q\mid i+r\mid j+s\mid k \qq 
(p, \; q, \; r, \; s \in {\cal H}) \q ,\] 
is still under investigation.

This work presents an explicit quaternionic translation of the standard 
(complex) theory based on the even group 
$SU(2, \; c)\times U(1, \; c)$. We wish to underline that while all even 
dimensional complex group can be translated into our quaternionic formalism 
with half the dimensions~\cite{del5} (an undoubted practical advantage), see 
$U(1, \; q)\mid U(1, \; c)$, work is still necessary for the translation of 
odd dimensional complex groups.

Having shown the possibility of rewriting standard theory in quaternionic 
form the question that comes to mind is if there exist interesting 
quaternionic equations corresponding to new physics. The analogy is always 
with the Schr\"odinger equation. This equation can always be written as a 
{\em pair} of real equations but the existence of complex numbers, for 
example in the wave function would be evidenced by the ``rule'' for 
expressing probability amplitudes in terms of the two real wave functions. 
In this context we recall the work of Adler~\cite{adl2} who assumes 
quaternionic probability amplitudes formulating a quite revolutionary 
quantum theory.

We conclude by hoping that the work presented in this paper be considered 
an encouragement for the use of a quaternionic quantum mechanics with complex 
geometry.

\bref
\bi{lam}
J.~Lambek, {\it If Hamilton had prevailed: Quaternions in Physics} 
(McGill University, Nov.~1994).
\bi{ham}
W.~R.~Hamilton, {\it Elements of Quaternions} (Chelsea, New York, 1969).
\bi{del1}
S.~De Leo, {\it Quaternions and Special Relativity}, J.~Math.~Phys. 
(to be published).
\bi{rot}
P.~Rotelli, \mxb{4}{933}{89}. 
\bi{hor}
L.~P.~Horwitz and L.~C.~Biedenharn, \axp{157}{432}{84}. 
\bi{adl}
S.~L.~Adler, {\it Quaternionic Quantum Mechanics and Quantum Fields} 
(Oxford, New York, 1995), p.~58-63.
\bi{del2}
S.~De Leo and P.~Rotelli, \pxf{45}{575}{92}.
\bi{del3}
S.~De Leo and P.~Rotelli, {\it Quaternion Higgs and the Electroweak Gauge 
Group}, Int.~J.~Mod.~Phys.~A (to be published).
\bi{del4}
S.~De Leo and P.~Rotelli, {\it Quaternionic Dirac Lagrangian}, 
Int.~J.~Mod.~Phys.~A (to be published).
\bi{itz}
C.~Itzykson and J.~B.~Zuber, {\it Quantum Field Theory} 
(McGraw-Hill, New York, 1985), p.~49.
\bi{del5}
S.~De Leo and P.~Rotelli, \pxxa{92}{917}{94}.
\bi{gil}
R.~Gilmore, {\it Lie Groups, Lie Algebras, and Some of Their Aplications} 
(John Wiley \& Sons, New York, 1974), p.~47.
\bi{adl2}
S.~L.~Adler,  \cxf{104}{611}{86}; \pxf{34}{1871}{86}; \xxx{37}{3654}{88}; 
\pxi{221B}{39}{89}; \xxx{332B}{358}{94}; \nxb{B415}{195}{94}.
\eref

\end{document}